\documentclass[final,3p,times,twocolumn]{elsarticle}

\usepackage{amssymb}
\usepackage{graphicx}% Include figure files
\usepackage{bm}% bold math
\usepackage{float}
\usepackage{xcolor}
\usepackage{etoolbox}
\usepackage{amsmath}
\biboptions{sort&compress}

\makeatletter
\def\ps@pprintTitle{%
\let\@oddhead\@empty
\let\@evenhead\@empty
\def\@oddfoot{}%
\let\@evenfoot\@oddfoot}

\def \beq {\begin{equation}}
\def \eeq {\end{equation}}

\begin{document}

\begin{frontmatter}

\title{Multi-photon enhancement of the Schwinger pair production \\ mechanism under strong FEL radiation}

\author{G. Mouloudakis and P. Lambropoulos}

\address{Department of Physics, University of Crete, P.O. Box 2208, GR-71003 Heraklion, Crete, Greece
\\
Institute of Electronic Structure and Laser, FORTH, P.O.Box 1527, GR-71110 Heraklion, Greece}

\begin{abstract}
In this work we study the production of electron-positron pairs from vacuum in presence of a field resulting from the collision of two counter-propagating FEL beams that undergo Gaussian amplitude fluctuations. Our work aims to the extension of previous works based on the standing wave hypothesis, by including the inherent stochastic amplitude fluctuations of the individual FEL beams. As shown, depending on the order of the process, a large non-linear enhancement in the number of created pairs can be expected over a big intensity window in the multi-photon regime. Vacuum pair creation in view of future plans on the production of ultra-strong and high energy radiation in FEL facilities is also discussed.
\end{abstract}

\end{frontmatter}

\section{Introduction}

The spontaneous production of electron-positron pairs in vacuum in the presence of ultrastrong electromagnetic radiation, is one of the most intriguing non-linear phenomena predicted by Quantum Electrodynamics (QED) theory, which  still has not been confirmed experimentally. The  analysis of this strong-field phenomenon was presented in 1931 in the work of Sauter \cite{ref1} while Schwinger was the first to calculate the rate of this process in the presence of a static, homogeneous electric field \cite{ref2}. According to \cite{ref2}, the spontaneous creation of an $e^{-}e^{+}$ pair in vacuum, often referred to as Schwinger mechanism, acquires a sizable rate for fields close to the critical value $\mathcal{E}_{c}=\frac{{m_{e}}^2c^3}{e\hbar}=1.32 \times 10^{16}$ V/cm. Due to the practical impossibility of creating such an ultrastrong static and homogeneous field in the laboratory, there was a lot of theoretical activity in the decade of 1970 on the analysis of vacuum pair creation using time-dependent electric fields \cite{ref3,ref4,ref5,ref6,ref7,ref8,ref9}. However, within the available optical laser technology at that time, the resulting rate would be far too small to be experimentally observable, due to insufficient power density necessary for breaking the vacuum.

The more recent advent of the free-electron laser (FEL), in facilities such as the European XFEL \cite{ref10}, DESY \cite{ref11}, SLAC \cite{ref12} and SACLA \cite{ref13}, revived the interest in this problem over the last twenty years or so \cite{ref14,ref15,ref16,ref17,ref18,ref19,ref20,ref21,ref22,ref23,ref24,ref25,ref26,ref27}, because FEL's can provide strong, tightly focused radiation with energies up to the hard X-rays regime. At the same time, many authors have in addition explored possible ways to effectively enhance the pair creation yield by using colliding focused polarized pulses \cite{ref28}, consecutive pulses with time delay \cite{ref29}, combinations of weak high-frequency and strong low-frequency fields \cite{ref30,ref31,ref32}, strong fields combined with thermal backgrounds \cite{ref33}, as well as fields with frequency \cite{ref34} or amplitude \cite{ref35} modulations. It is important to note here that most of the  above mentioned studies distinguish pair creation into two regimes depending on the field parameters, i.e. the tunneling and the multi-photon regime. For a harmonically oscillating field $E (t) =\mathcal{E}cos(\omega t)$, these two regimes are reflected in the value of $\gamma=\frac{m_{e} c \omega}{e \mathcal{E}}$, which is the exact analog of the Keldysh parameter in strong-field laser-atom interactions. In particular, for $\gamma \ll 1$ (high-field, low-frequency limit) $e^{-}e^{+}$ pairs are mainly created via the tunneling effect, whereas for $\gamma \gg 1$ (low-field, high frequency limit) the pairs are created via multi-photon absorption with the corresponding formulas resembling a perturbative result. Pair creation in the intermediate regime between those two, often referred as non-perturbative multi-photon regime, has also been considered in a number of papers \cite{ref36,ref37,ref38,ref39,ref40}. It must always be kept in mind that the notion of strong non-perturbative regime depends not only on the intensity but also on the wavelength. For example, infrared radiation of intensity $10^{14}$ W/$cm^{2}$ induces non-perturbative behavior, whereas the same intensity at XUV belongs to the multi-photon regime \cite{ref41}. 

Although pair creation via the Schwinger mechanism has not been experimentally observed so far, creation of $e^{-}e^{+}$ pairs via the so-called Breit-Wheeler mechanism \cite{ref42} has been reported since the mid 1990s in pioneering experiments at Stanford (SLAC) \cite{ref43,ref44}. In those experiments, a high-energy photon created through Compton backscattering of optical photons by a 46.6 GeV electron beam, collided with laser photons of wavelength 527nm to produce electron-positron pairs. The reported measured signal of 106 $\pm$ 14 positrons were the first data involving the creation of $e^{-}e^{+}$ pairs in the laboratory. The theoretical analysis of SLAC's data was based on the existing non-perturbative theory of the multi-photon Breit-Wheeler reaction available at the time \cite{ref45}, whereas further theoretical insights of the measurements where added many years later \cite{ref46}.

Given the huge theoretical activity over the past few years on the topic of vacuum pair creation using ultrastrong radiation \cite{ref47}, it is the purpose of this paper to introduce and explore an aspect that seems to have escaped attention until now. That is the possibility of enhancement of the Schwinger mechanism  owing to the intrinsic stochastic fluctuations of FEL sources, and if yes, over which range of intensities such enhancement could be expected to be significant. The stochastic properties of self-amplified spontaneous emission (SASE) FEL sources have been documented quite extensively through theoretical \cite {ref48,ref49,ref50,ref51,ref52,ref53} as well as experimental studies \cite{ref54,ref55,ref56,ref57,ref58,ref59}. Briefly, their stochastic properties are akin to those of chaotic (thermal) radiation \cite{ref60}. The feature of that radiation of direct relevance to our considerations is the strong intensity fluctuations, implying spiky behavior during the pulse, which is known to affect the yield of processes whose rate depends on the intensity in a non-linear fashion \cite{ref50,ref51,ref61}.

Before embarking on the formal and computational details of the problem, it would be helpful to establish the qualitative context of the problem at hand. In any arrangement involving a laser for the observation of pair creation, that laser will necessarily be pulsed. The particular temporal shape is of secondary importance but the pulse duration, as characterized by the full width at half maximum (FWHM), would be of decisive importance. To clarify the reason for its importance we need to consider the various time scales at play in the underlying physical process. One of those time scales is the transition time (TT), which is the time within which the photons must be absorbed for the process to be completed. The initial state is the vacuum with the final state being the continuum of the pair $e^{-}e^{+}$. The energy gap that the absorption of the photons must bridge for the pair to be created is $1.022 MeV$. There are no real intermediate states within the gap. The process has to proceed via virtual intermediate states of the continuum which energetically lie $1.022 MeV$ above the initial state. Let us take, for example, the case of 10 photon absorption which would require photons of energy a bit more than $0.1 MeV$, which is also the energy of the first virtual state. Its wavefunction consists of a linear superposition of all continuum states. Its "lifetime" is the inverse of its detuning from the nearest  state with allowed transition, which in this case is the continuum threshold detuned by about $0.9 MeV$. This corresponds to a lifetime of about $4\times10^{-6}  fs$. Obviously the process must be completed within that time. Put otherwise, the other photons must arrive before the virtual state ceases to exist. The laser pulse duration provides the second time scale, the interaction time (IT), by which we mean the time interval during which the fields are present to induce pair creation. For the time being and possibly the near future, the pulse duration of any conceivable laser at any wavelength will be orders of magnitude longer than the above TT. Therefore, the IT can be assumed to be much longer than the TT. If we are to contemplate a short wavelength laser in the X-ray range, even for photon energy of say $10 keV$, the field period (the third time scale) is about $4\times10^{-4}  fs$, two orders of magnitude longer than the TT. For all practical purposes, at present the pulse duration of an X-ray FEL cannot be expected to be much shorter than $0.1-1  fs$. In the light of the above time scales, it should be clear that the TT is so short compared to the other time scales that pair creation in the presence of a laser is essentially an instantaneous process. 

Let us now examine what exactly  happens during a pulse. Assume  that the intensity is such that at the peak of the pulse $\gamma\ll 1$ which entails tunneling. As the pulse rises and falls, however, it goes through ranges of intensity for which $\gamma\gg 1$ where multi-photon transition dominates and which actually is more efficient \cite{ref5,ref6,ref15}. This means that under the  reality of a pulse, in the end the pairs produced will have been created via multi-photon absorption, as well as tunneling around the peak of the pulse. If the temporal pulse shape is smooth, say a Gaussian, all we need is to integrate the pair production over the pulse shape. If on the other hand the pulse is not smooth but the intensity undergoes stochastic fluctuations, integration over a pulse shape is impossible. In that case, we need to work with averaging over the stochastic fluctuations. If the  equations governing the rate of the process depend on the laser intensity non-linearly, the averaging does not lead to the replacement of the instantaneous intensity by the average intensity, as would have been  the case for a linear dependence. It does instead lead to a more complex dependence, the form of which depends on the stochastic properties of the source and the type of non-linearity. The possible relevance of that issue stems from a well established precedence in the long history of multi-photon processes in atomic systems. It is known that the rate of an $N$-photon process, far from resonance  with real intermediate states, depends on the $Nth$ order intensity correlation function \cite{ref61,ref62,ref63,ref64}. And since correlation functions depend on the stochastic properties of the radiation, the rate of even the simplest non-linear process such as the above can be affected dramatically by intensity fluctuations. If we consider, for example, an 11-photon process induced by thermal radiation whose $Nth$ order intensity correlation function is given by $N!\times I^{N}$, where $I$ is the average intensity, that process would be enhanced by a factor of about $10^{7}$, as  compared to the same process under a smooth pulse without the fluctuations of the thermal source. That is an example of a relatively simple nonlinearity, in that the dependence on intensity is proportional to a single correlation function, which reduces to a power of the intensity for a smooth pulse.

The situation becomes much more complex when for a smooth pulse the rate is a complicated function of the intensity, as that entails the dependence of the rate on a complicated function of a stochastic variable. In that case, it is far from evident whether averaging over the fluctuations will lead to enhancement of the rate. One might expect that in the regime of multi-photon dominance in the rate, there may be some enhancement, the details of which are examined in the sections that follow.  A helpful physical picture provides some insight into the role of intensity fluctuations, if we note that intensity fluctuations such as those of thermal radiation imply random spikes of intensity during the pulse. Some of those spikes will have a peak intensity much higher than the peak of the average intensity. Since a non-linear function of the intensity involves high powers of the instantaneous values, spikes with intensity higher than the average will tend to enhance disproportionally the high powers of the function. In the case of a process depending on a single power of the intensity, the enhancement caused by spikes is obvious. To summarise our basic argument, the time scales entering pair production mediated by laser radiation satisfy the conditions for the influence of intensity fluctuations on the rate of the process. 

The case of short wavelength FEL is of particular interest in that context, because as already noted above, typical FEL's are known to exhibit strong intensity fluctuations very similar to those of thermal (chaotic) radiation. Although, in principle, the effect of fluctuations can be analyzed mathematically for any photon frequency, in this paper we restrict our analysis to very short wavelength FEL sources of photon energies above the $10 keV$ range. For the time being, pulse durations of FEL's are at best in the fs range, with plans  underway towards sub-femtosecond pulses. Even if we were to assume pulse duration of 1 attosecond, in the photon energy range of $10-100 keV$ such a pulse would span tens  of cycles, which makes it much larger than the TT. If the pulse udergoes intensity fluctuations, the relative time scales allow their influence on the rate of pair creation. It bears repeating that whether the result is enhancement of the rate, remains to be explored through the quantitative analysis in the sections that follow. 

Aside from the theoretical ideas on the possible role of intensity fluctuations in pair creation, our motivation for this work was inspired by known results in multi-photon ionization of atoms, where enhancement of the order of $N!$ has been observed a long time ago, in processes of order as high as $N=11$ \cite{ref61}. In atoms, in addition to the initial ground state and the ionization potential, there are infinitely many real intermediate states which can happen to be in near resonance with the absorption of a number of photons smaller than $N$. When the order of the overall process is, let us say 10, at most 2 but most often 1 real intermediate states can be in near resonance to a degree that can violate the overall non-resonant condition. In that case, the enhancement will be slightly lower than $N!$, but still quite significant, as large as for example a factor  $10^{7}$ observed in \cite{ref61}.  Curiously, there is one atomic system whose ionization bears an uncanny similarity with pair creation. That is the negative ion of hydrogen whose binding energy is $0.754 eV$. It has a ground state and a continuum, without any real intermediate states. Although at much lower photon energies, its multi-photon ionization (technically called detachment) has been studied theoretically as well as experimentally \cite{ref65,ref66}. There is on the other hand a crucial difference between multi-photon break up of an atom or molecule and pair creation in vacuum. The interaction region in laser-atom interactions contains a finite number of atoms. As a result, even unlimited intensity will not help beyond the point of complete ionization of the species in the interaction volume. And this can happen even before the pulse reaches its peak \cite{ref67}. Vacuum on the other hand is an infinite sea from which pairs can be created as long as there are photons present. It is therefore reasonable to expect that, at least in the multi-photon regime of $\gamma\gg 1$ an enhancement of the order of $N!$ would be obtained. As already noted above, beyond that regime of intensities, novel behavior should be expected and indeed obtained through our calculations. The discussion of the feasibility of experiments compatible with our calculations is deferred until the last section of conclusions.

The article is organized as follows: In Sec. 2 we provide a brief discussion about the conservation laws of Schwinger pair creation and examine the problem of pair creation under the standing wave hypothesis in the presence of field amplitude fluctuations. In Sec. 3 we analytically prove that the presence of such amplitude fluctuations can lead to a non-linear multi-photon enhancement of vacuum pair creation in the zero field limit and show numerically that this enhancement persists over a big intensity window that depends on the order of the process. In Sec. 4 we explore the possibility of observing pair creation in view of future plans on the production of ultra-strong and high energy radiation in FEL facilities, while in Sec. 5 we provide some concluding remarks.

\section{Conservation laws and the standing wave hypothesis}

Electromagnetic fields can be characterized using a gauge and Lorentz invariant approach in terms of the electromagnetic field strength tensor $F^{\mu \nu}$ and its dual, ${\tilde F^{\mu \nu }} = {\frac{1}{2}}{{\epsilon}^{\mu \nu \alpha \beta }}{F_{\alpha \beta }}$ (${{\epsilon}^{\mu \nu \alpha \beta }}$ is the rank-four Levi-Civita tensor), as:
\begin{subequations}
\begin{align}
\mathcal{F} = {\frac{1}{4}}{F_{\mu \nu }}{F^{\mu \nu }} =- {\frac{1}{2}}({{\vec E}^2} - {{\vec B}^2})\\
\mathcal{G} = {\frac{1}{4}}{F_{\mu \nu }}{{\tilde F}^{\mu \nu }} = c\vec E \cdot \vec B
\end{align}
\end{subequations}
It is generally known that $e^{-}e^{+}$ pair creation cannot occur either in the light-like field of a plane monochromatic wave, which is always characterized by $\mathcal{F}=0=\mathcal{G}$, or in a field characterized by $\mathcal{F}>0$ and $\mathcal{G}=0$, corresponding to a pure magnetic field. On the other hand, pair creation can occur in fields described by $\mathcal{F}<0$ and $\mathcal{G}=0$. Brezin and Itzykson were the first to calculate the pair production rate by a spatially uniform periodic electric field \cite{ref4}, stating that such a field could by realised by proper optical focusing of laser beams in the laboratory. In view of some concern about pair production using this method \cite{ref68}, it was argued that another more effective method of realising such a field would be via the creation of a standing wave formed by the superposition of two counter-propagating coherent laser beams of the same wavelength $\lambda$ \cite{ref5,ref6,ref8,ref9,ref16,ref24,ref36}. In this idealized scenario pair production can occur in an antinode of the standing wave at lengths $l \ll \lambda$, such that the spatial inhomogeneities of the field can be neglected to a good approximation. A detailed analysis of the Schwinger pair creation using counter-propagating laser pulses can be found in the paper by G. R. Mocken et al. \cite{ref37}. We should also note that pair production can occur in plasma-like media as has been discussed in a series of papers by H. K. Avetissian et al. \cite{ref69,ref70,ref71,ref72}.

Following the lines of previous works based on the conceptual experiment of pair creation by an idealized standing wave, we generalize part of the problem by including the presence of amplitude (intensity) fluctuations in the initial beams and show that such fluctuations could eventually lead to large enhancement of the created pairs. Specifically, we focus on the study of pair creation resulting from a standing wave formed by the interference of two counter-propagating FEL beams of the same wavelength that undergo amplitude fluctuations, i.e. beams of the form $\mathcal{E}_{t}^{(1)}cos(\omega t - \textbf{k}\cdot\textbf{r})$ and $\mathcal{E}_{t}^{(2)}cos(\omega t + \textbf{k}\cdot\textbf{r})$, where the index ``t'' denotes the stochastic character of the amplitudes and \textbf{k} is the beam wavevector. In this scenario, the resulting field is equal to $(\mathcal{E}_{t}^{(1)}+\mathcal{E}_{t}^{(2)})cos(\omega t)cos(\textbf{k}\cdot\textbf{r})+(\mathcal{E}_{t}^{(1)}-\mathcal{E}_{t}^{(2)})sin(\omega t)sin(\textbf{k}\cdot\textbf{r})$. Adopting the approximation $\textbf{k}\cdot\textbf{r} \ll 1$, which is based on the assumption that pair creation occurs at lengths smaller than the wavelength of the beams \cite{ref36}, the resulting field is, to a good approximation, a standing wave of the form $\mathcal{E}_{t}cos(\omega t)$, where $\mathcal{E}_{t}=\mathcal{E}_{t}^{(1)}+\mathcal{E}_{t}^{(2)}$.
In particular, we consider the case in which both $\mathcal{E}_{t}^{(1)}$ and $\mathcal{E}_{t}^{(2)}$ undergo Gaussian amplitude fluctuations, corresponding to those of an ideal chaotic state. In that case, the first-order intensity correlation function $\tilde G_1$ of the resulting field is two times the first-order intensity correlation function of the chaotic field $G_1^{chao}$, since:
\beq
\begin{split}
\tilde{G_1} & \equiv \left\langle {\mathcal{E}_t^*{\mathcal{E}_{t'}}} \right\rangle = \left\langle {\left( {\mathcal{E}{{_t^{(1)}}^*} + \mathcal{E}{{_t^{(2)}}^*}} \right)\left( {\mathcal{E}_{t'}^{(1)} + \mathcal{E}_{t'}^{(2)}} \right)} \right\rangle \\ 
& =\left\langle {\mathcal{E}{{_t^{(1)}}^*}\mathcal{E}_{t'}^{(1)} + \mathcal{E}{{_t^{(1)}}^*}\mathcal{E}_{t'}^{(2)} + \mathcal{E}{{_t^{(2)}}^*}\mathcal{E}_{t'}^{(1)} + \mathcal{E}{{_t^{(2)}}^*}\mathcal{E}_{t'}^{(2)}} \right\rangle \\
& = \left\langle {\mathcal{E}{{_t^{(1)}}^*}\mathcal{E}_{t'}^{(1)}} \right\rangle  + \left\langle {\mathcal{E}{{_t^{(2)}}^*}\mathcal{E}_{t'}^{(2)}} \right\rangle  = 2{G_1^{chao}} ,
\end{split}
\eeq
given the fact that no amplitude correlation exists between the two beams. It is straightforward to show that this result persists even for correlations functions of arbitrary order. Therefore, for the Nth-order (normal-ordered) intensity correlation function of the resulting field  we can prove that
\beq
\tilde G_N = 2 G_N^{chao}
\eeq
The effects of this result on pair creation are discussed in the next section. A brief clarification should be made at this point regarding  the connection with FEL radiation: As is generally known, apart from amplitude fluctuations, FEL radiation does also exhibit fluctuations in phase \cite{ref48,ref50,ref51}, making the standing wave hypothesis \cite{ref5,ref6,ref8,ref9,ref16,ref24,ref36} even more challenging to realise experimentally. The reason we do not include this type of fluctuations in the formulation is twofold: First, the problem of vacuum $e^{-}e^{+}$ pair creation in the presence of fields that undergo fluctuations both in amplitude and phase is considerably more complex, while the tools that have been developed for treating analogous problems in atomic transitions \cite{ref62,ref63,ref64} are not directly applicable to the problem at hand. Second, the inclusion of phase fluctuations resulting to a finite field bandwidth is not expected to notably affect the measured signal of created pairs, since the process of pair creation, when operating in the multi-photon regime, is ultimately an N-photon escape process, involving no intermediate resonances. And in this case the role of field bandwidth is essentially irrelevant. Therefore, in this paper we are interested solely in examining how the presence of fluctuations in the amplitude of the resulting standing wave may affect the total number of  created pairs.

\section{Multi-photon enhancement of vacuum pair creation}

A few years after the paper of Brezin and Itzykson \cite{ref4}, Popov extended their work using the imaginary time method \cite{ref73} and determined more accurately the pre-exponential factor in the expression of the rate of vacuum pair creation. For the details of this derivation we refer the reader to \cite{ref5,ref6,ref15,ref27,ref74}. Assuming a monochromatic laser field of the form $\mathcal{E}cos(\omega t)$ focused down to the diffraction limit, Popov obtained the following expressions for the number of $e^{-}e^{+}$ created pairs in the $\gamma \ll 1$ and $\gamma \gg 1$ limits, corresponding to the tunneling and multi-photon regimes, respectively:
\beq
\begin{split}
& {N(\mathcal{E})}={2^{-3/2}}N_0^4{{\left(\frac{\mathcal{E}}{\mathcal{E}_c}\right)}^{5/2}}\\
& \times \exp\left[ {-\frac{\pi \mathcal{E}_c}{\mathcal{E}}\left( {1 - {\frac{1}{{2{{{N_0}^2\left( {{\frac{\mathcal{E}}{\mathcal{E}_c}}} \right)}^2}}} }} \right)} \right]\left(\frac{\omega \tau }{2\pi}\right)
\quad\mathrm{,}\quad \gamma \ll 1 
\end{split}
\eeq
\beq
{{N(\mathcal{E})} \approx 2\pi N_0^{3/2}{{\left({\frac{8{\mathcal{E}_c}}{{N_0}e\mathcal{E}}}\right)}^{ - 2{N_0}}}\left(\frac{\omega \tau }{2\pi}\right)}
\quad\mathrm{,}\quad
\gamma \gg 1
\eeq
where $\mathcal{E}$ is the amplitude of the electric field, $\mathcal{E}_{c}=\frac{{m_{e}}^2c^3}{e\hbar}=1.32 \times 10^{16}$ V/cm is the critical electric field value calculated by Schwinger \cite{ref2}, $\tau$ is the interaction time and $N_0=\frac{2mc^2}{\hbar \omega}$ is the minimum number of photons needed for vacuum pair creation to occur at a given frequency. Comparison between Eqs. (4) and (5) reveals that pair creation in the multi-photon regime is far more effective than pair creation in the tunneling regime due to the exponential suppression of the latter for $\mathcal{E}<\mathcal{E}_c$. Given that the electric field amplitude is proportional to the square root of the intensity, equations (4) and (5) can also be written in the form:
\beq
\begin{split}
& {N(I)} = {2^{ - 3/2}}N_0^4{{\left(\frac{I}{I_c}\right)}^{5/4}}\\
& \times \exp\left[ {-\pi\sqrt{\frac{ I_c}{I}}\left( {1 - {\frac{1}{{2{N_0}^2{{\left( {{\frac{I}{I_c}}} \right)}}}} }} \right)} \right]\left(\frac{\omega \tau }{2\pi}\right)
\quad\mathrm{,}\quad
\gamma \ll 1
\end{split}
\eeq
\beq
{{N(I)} \approx 2\pi N_0^{3/2}{{\left({\frac{8}{{N_0}e}}\right)}^{ - 2{N_0}}}{\left(\frac{I}{I_c}\right)}^{N_0}\left(\frac{\omega \tau }{2\pi}\right)}
\quad\mathrm{,}\quad
\gamma \gg 1
\eeq
where $e$ is the Euler's constant, $I$ is the intensity of the electric field and $I_{c}= 4.65 \times 10^{29}$ W/cm$^{2}$ is the intensity corresponding to $\mathcal{E}_{c}$.

One of the key results in quantum optics when working under the zero bandwidth approximation is that the effects of any type of radiation field on a transition can be captured by first solving the problem assuming a harmonically oscillating field of constant amplitude ($\mathcal{E}cos(\omega t)$) and then average over the intensity distribution corresponding to the radiation field considered \cite{ref62}. We will use this method to calculate the number of pairs created by the amplitude fluctuating standing wave field considered in the previous section.

Inspection of Eq. (7) reveals that the number of pairs created for $\gamma \gg 1$ is proportional to the intensity to the power of $N_0$, i.e. the $N_0$-th order intensity correlation function of the harmonically oscillating field of constant amplitude, as expected for a multi-photon transition that involves no intermediate resonances. Given that: (i) all the information about the effects of the coherence properties of a considered field on such a multi-photon transition is contained in its intensity correlation function, (ii) the $N$-th order intensity correlation function of the standing wave field resulting from the interference of the two counter-propagating beams with Gaussian amplitude fluctuations is twice that of the chaotic field as shown in the previous section and (iii) the intensity distribution of the chaotic field is $p(I')=\frac{e^{-\frac{I'}{I}}}{I}$, where I is the mean intensity, it is straightforward to argue that the resulting number of $e^{-}e^{+}$ pairs created by such a standing wave field, in the zero bandwidth approximation, is:
\beq
\tilde N (I)=2\int_{0}^{\infty}N\left(I'\right)\frac{e^{-\frac{I'}{I}}}{I}dI'
\eeq
where the factor of 2 is direct consequence of Eq. (3). 

Using now the expressions for the number of created pairs in each regime, one can show that in the limit of zero intensities, the number of $e^{-}e^{+}$ pairs created by such a field is $2N_0!$ times larger than the respective number of pairs created in the case of a harmonically oscillating of constant amplitude, i.e. 
\beq
\mathop {\lim }\limits_{I \to {0^+}} \frac{\tilde N (I)}{N(I)} = 2N_0!
\eeq
where $N_0$ is the minimum number of photons participating in the process. The details of this calculation is presented in the appendix. Note that since $\gamma$ is inversely proportional to the square root of the intensity and the integral over the intensity in Eq. (8) extends from zero up to infinity, we split the integration into two parts; namely from 0 to $I_b$ (multi-photon regime) and from $I_b$ up to infinity (tunneling regime), where $I_b$ is the intensity corresponding to $\gamma=1$. We should note that this method is not generally valid for arbitrary intensities since (i) there practically exists no sharp boundary between these two regimes and (ii) the considered formulas (6) and (7) give only an approximate expression of the number of created pairs in the $\gamma \sim 1$ regime in which they are joined in order of magnitude as $N_0$ decreases. However, in what follows we will limit our discussion to intensities lower than $I=0.01 I_c$. In this case the considered approximation is expected to give a good estimation of the total number of created pairs and the reason behind this has to do with the specific form of the intensity distribution $p(I')=\frac{e^{-\frac{I'}{I}}}{I}$ as a function of $I'$ for different values of $I$. In particular, $p(I')$ is decreasing for increasing $I'$ but this occurs faster for smaller values of I. And even in the maximum considered case of $I=0.01 I_c$, due to the exponential factor, the contribution of the terms with about $I' \geq 0.02 I_c$ is already small. This has two consequences: Firstly, the integral corresponding to the tunneling regime, i.e. $2\int_{I_b}^{\infty}N\left(I'\right)\frac{e^{-\frac{I'}{I}}}{I}dI'$ does not give a substantial contribution to the total number of created pairs $\tilde N$. This is due to the fact that the boundary intensity corresponding to $\gamma =1$, which can be expressed as $I_b=\frac{4}{N_0^2}I_c$, is already larger than $0.02 I_c$ for all $N_0$ up to 14, resulting to the suppression of the number of pairs due to the exponentially decaying factor. Therefore the main contribution to the total number of pairs generally arises from the multi-photon integral $2\int_{0}^{I_b}N\left(I'\right)\frac{e^{-\frac{I'}{I}}}{I}dI'$. Secondly, as far as the multi-photon integral is concerned, even if Eq. (7) gives only an order of magnitude approximation of the created pairs as we approach the $\gamma \sim 1$ regime, or in other words the intensity boundary $I_b$, if the considered number of photons participating in the process is approximately up to 14, the terms in the vicinity of the non-perturbative multi-photon regime ($\gamma \sim 1$) have relatively smaller weights and do not contribute significantly to the resulting number of created pairs $\tilde N$. In view of the above, we limit our discussion to photon orders up to about $N_0=14$ and to intensities up to about $I=0.01I_c$ in order to ensure the validity of our approximation. Note that for arbitrary intensities and orders $N_0$ one cannot use this approach to the problem at hand.

Using this method, one can prove the validity of Eq. (9) in the zero field limit, regardless of the exact choice of the boundary (this is due to the consequence of taking the limit  $I \to {0^ + }$ as discussed in the appendix). 

\begin{figure}[H] 
	\centering
	\includegraphics[width=7.8cm]{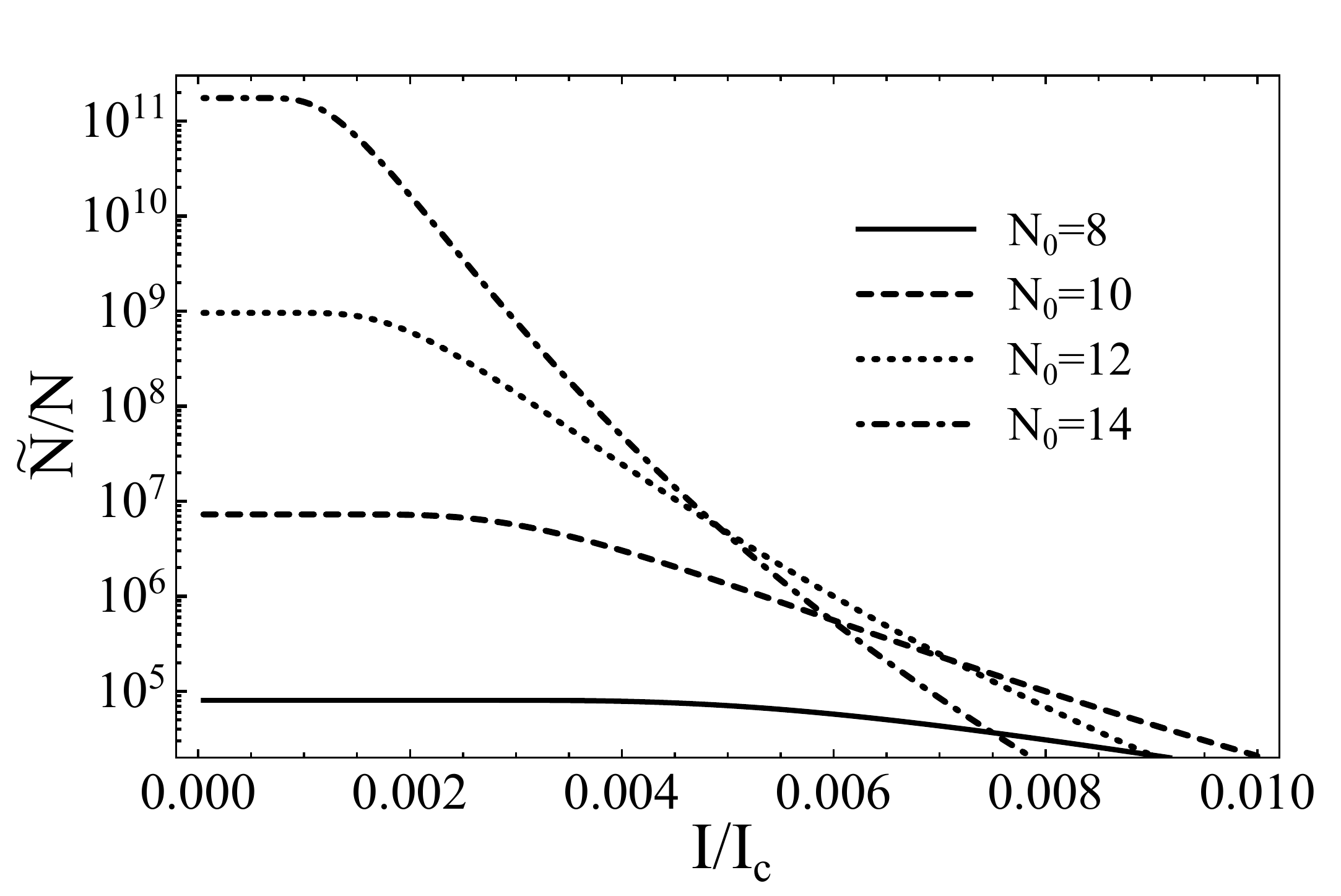}
		\caption[Fig.1]{Ratio of the number of $e^{-}e^{+}$ pairs created by a field that results from the collision of two counter-propagating FEL beams that undergo Gaussian amplitude fluctuations over the number of $e^{-}e^{+}$ pairs created by a harmonically oscillating field of the form $\mathcal{E}cos(\omega t)$. The ratio is plotted as a function of the intensity for different values of $N_0$. The range of the intensity plateau over which the ratio remains equal to predicted low field value 2$N_0$! is increasingly larger for smaller values of $N_0$. The values of $N_0$ used are: $N_0=8$ (solid line), $N_0=10$ (dashed line), $N_0=12$ (dotted line) and $N_0=14$ (dash-dotted line).}
\end{figure}

Eq. (9) is a direct consequence of the proportionality of the number of created pairs to the intensity correlation function of the field, which for a chaotic field is $N_0!$ larger than the respective correlation function of the coherent field. Of course in the limit of zero intensities the number of created pairs is practically zero, therefore the presence of such an enhancement factor does not seem to be useful. However, numerical calculations of the created pairs number ratio as a function of the intensity using Eqn. (8), reveal that this enhancement factor persists over a large intensity window which depends on the order of the process, extending up to the strong field regime (Fig. 1).

 In Fig.1 we plot the ratio of the number of $e^{-}e^{+}$ pairs created by the field resulting from the collision of two counter-propagating FEL beams that undergo amplitude fluctuations with Gaussian statistics over the number of $e^{-}e^{+}$ pairs created by a harmonically oscillating field, for various orders $N_0$. As seen, there exists a large intensity plateau over which the ratio remains equal to the expected value $2N_0$! of the zero field limit. The range of the intensity plateau strongly depends on the order of the process $N_0$, as it decreases while $N_0$ is increased. Given that the enhancement factor 2$N_0$!, depending on the minimum number of photons participating in the process, can reach very high values, it is the purpose of the next section to examine  whether and under which combination of parameters such an enhancement could be utilized to experimentally observe the Schwinger vacuum pair creation under the considered configuration.

\section{Results and Discussion}
Given that the parameter $\gamma$ is expressed in terms of the intensity as $\gamma = \frac{\hbar\omega}{mc^2}\sqrt{\frac{I_{c}}{I}}$ and that the minimum number of photons participating in the process is $N_0 = \frac{2mc^2}{\hbar\omega}$, we can write that: 
\beq
\gamma = \frac{2}{N_0}\sqrt{\frac{I_c}{I}}
\eeq
According to Eq. (10), $\gamma$ is inversely proportional to the order of the process $N_0$. Let us for example examine the case of $N_0 =14$, corresponding to the photon energy $\hbar\omega = 73 keV$. In this case the enhancement factor $\tilde N / N$ reaches the maximum value 2$N_0$!$ = 1.74 \times 10^{11}$ and persists for intensities up to $I \cong 0.001 I_c$ (see Fig.1). However, even if the enhancement factor seems very large, for intensities around $I \cong 0.001 I_c$ the number of created pairs by an harmonically oscillating field is of the order of $10^{-15}$ (for an interaction time $\tau =50fs$), therefore an enhancement factor of the order of $10^{11}$ does not lead to an observable number of pairs. In this case one could try to increase the intensity in order to create more pairs. However, the drawback in this case would be that, for the intensities such that the number of created pairs is measurable, according to Eq. (10), the parameter $\gamma$ would be such that pair creation would gradually approach the tunneling regime. On the other hand, using a much lower photon number, say $N_0=5$, the maximum enhancement factor would be only 2$N_0$!$=240$ and the corresponding energy of the photons needed to observe the process would be extremely high ($\hbar\omega \approx 204 keV$). 

\begin{figure}[H] 
	\centering
	\includegraphics[width=8cm]{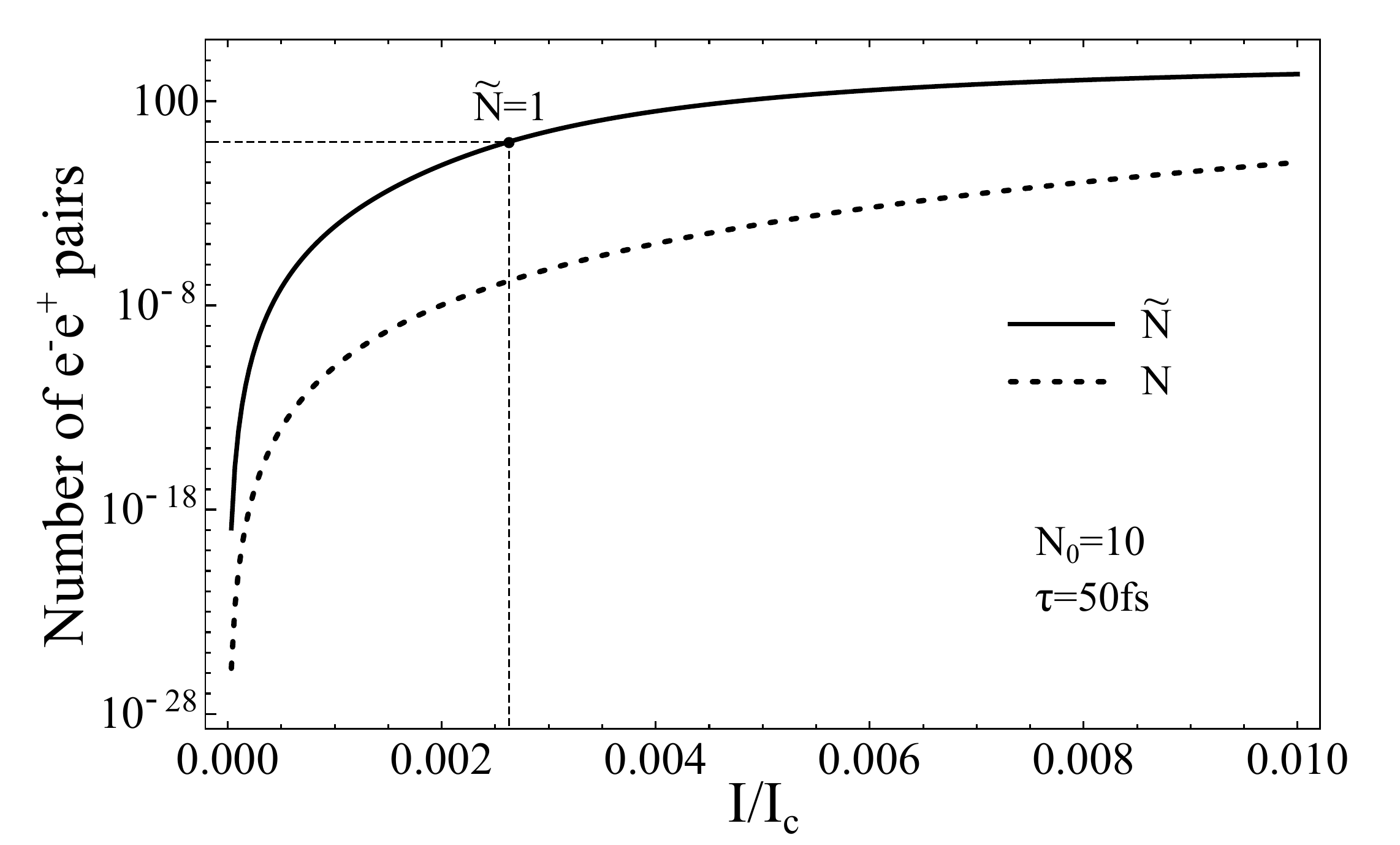}
		\caption[Fig.2]{Number of $e^{-}e^{+}$ pairs created by a field that results from the collision of two counter-propagating FEL beams that undergo Gaussian amplitude fluctuations (solid line), compared to the respective number of pairs created by a harmonically oscillating field of the form $\mathcal{E}cos(\omega t)$ (dotted line). The parameters used are: minimum number of photons $N_0 =10$ and interaction time $\tau = 50fs$. The vertical axis is presented in logarithmic scale and the dashed line corresponds to the intensity such that $\tilde N = 1$.}
\end{figure}

In view of the above, we consider an intermediate scenario where $N_0 = 10$, corresponding to photon energies $\hbar\omega \approx 102 keV$. The threshold of 100keV photons, even seemingly high, is expected to be reached in the near future based on recent studies on technical design plans and simulations confirming the possibility of obtaining such high-energy FEL photons in the European XFEL facilities \cite{ref75,ref76,ref77}. In addition to this, several years ago, works on multi-photon ionization experiments in atomic systems with chaotic radiation have reported the observation of the expected non-linear multi-photon enhancement for processes of such a high-order \cite{ref61}. Therefore it seems legitimate to investigate the case of pair creation via the absorption of 10 photons.

As can be seen in Fig. 1, for a 10-photon process,  a maximum enhancement factor of 2$N_0$!$=7.26 \times 10^6$, for intensities up to $I \cong 0.0028 I_c$ can be expected.
In Fig. 2 we plot the actual number of the created $e^{-}e^{+}$ pairs as a function of the intensity, both for the harmonically oscillating field of the form $\mathcal{E}cos(\omega t)$ (dotted line), as well as for the field resulting from the collision of two counter-propagating FEL beams that undergo amplitude fluctuations with Gaussian statistics (solid line). The calculation has assumed interaction time of $\tau =50fs$, but  one can easily obtain the resulting number of pairs for any interaction time, owing to the linear dependence of $N(I)$ and $\tilde N (I)$ on $\tau$. As is evident in Fig. 2, the number of pairs created using a field of the form $\mathcal{E}cos(\omega t)$ is very low for intensities below the critical intensity $I_c$. However, by taking advantage of the large enhancement factor for $N_0=10$, $\tilde N$ becomes sufficiently large to be experimentally detectable at lower intensities. The intensity at which $\tilde N =1$ (dashed line), is approximately $I \cong 0.0026 I_c$ and according to Eq. (10) corresponds to $\gamma \cong 3.9$. This intensity may become even lower as long as one can achieve larger interaction times $\tau$. Intensities of that magnitude are likely to be available in the future in FEL facilities such as the European XFEL \cite{ref75,ref76,ref77} or the SULF in China \cite{ref78}.

A few further comments regarding the assumptions underlying  this work are in order at this point. The derivation of Eqs. (4) and (5) for the number of created pairs provided by Popov are based on the assumption that the number of pairs are calculated in a volume equal to $\Delta V = {\lambda}^3$. This assumption can of course be revised by multiplying the resulting equations with the proper scaling factors. Note that one does not need to focus the radiation field down to the diffraction limit (which for high energy photons becomes increasingly difficult) in order to achieve the desired field intensity, as long as there is adequate power to balance the effect of less tight focusing. Moreover, even if the calculations are based on the standing wave hypothesis, extended in a way that includes the stochastic character of the amplitudes of the individual beams, it is obvious that ones does not need a standing wave in order to observe the non-linear multi-photon enhancement in pair creation induced by the fluctuating fields. The adoption of this hypothesis is useful on the one hand because of its analytical advantage and on the other hand because of the additional factor of 2 that one obtains according to Eq. (3). One could of course approach the problem from a solely numerical perspective, by using the quantum Vlasov equations along the lines of the work of I. Sitiwaldi and Bai-Song Xie \cite{ref35}, in combination  with a Gaussian stochastic modulation of the amplitude, using Monte Carlo techniques. However, this method would not provide any additional insight to the present work as far as the non-linear multi-photon enhancement factor is concerned. Finally, as described in the previous section, our calculations involve the notion of an intensity boundary between the multi-photon and the tunneling regime, necessary in order to perform the integration in Eq. (8). Even though  a sharp boundary does not really exist between these two regimes, splitting Eq. (8) into two integrals and calculating them separately using Eqs. (6) and (7) is expected to give a good estimate of the expected number of created pairs for $N_0=10$ and for intensities up to $I=0.01 I_c$, as discussed in section 3. 

\section{Concluding Remarks}
In the light  of our results, it appears that the role of intensity fluctuations in enhancing  non-linear strong field phenomena may be exploited advantageously in  Schwinger pair production, Our results, based as they are on approaches in the existing literature cited in our Introduction, may also be used in combination with other  methods aiming at  enhancing the yield of pair creation. To the best of our knowledge, the problem has so far been explored only theoretically. The stumbling block towards experimental verification has been the availability of appropriate laser sources. This is particularly crucial for FEL sources which, however, are still under constant development. The  source ideally suited for the experimental exploration of our predictions would be a SASE FEL with photon energy range around 100 keV. The possibility of developing FEL's in that photon energy range has in fact been addressed in quite recent theoretical studies \cite{ref75,ref76,ref77}. Considering the fact that about 40 years ago an X-ray laser was viewed as science fiction, in view of the rapid development of hard X-ray FEL's within the last 15 years or so, the availability in the near future of an FEL in the above photon energy range would not seem too optimistic. Be that as it may, intensity fluctuations are apt to play an intriguing role in pair creation under FEL radiation, if other issues related to focusing etc. can be addressed. 

\section*{Acknowledgments}
 GM would like to acknowledge the Institute of Electronic Structure and Laser (IESL), FORTH, for financially supporting this research by its Research and Development Programs.
 
\appendix
\section{}

In this appendix we prove that in the zero intensity limit, the number of $e^{-}e^{+}$ pairs created by the standing wave field that results from the collision of two counter-propagating FEL beams with Gaussian amplitude fluctuations, is $2N_0!$ times larger than the respective number of pairs created by a field of the form $\mathcal{E}cos(\omega t)$, i.e.
\beq
\mathop {\lim }\limits_{I \to {0^+}} \frac{\tilde N(I)}{N(I)} = 2N_0!
\eeq
where 
\beq
\tilde N (I)=2\int_{0}^{\infty}N\left(I'\right)\frac{e^{-\frac{I'}{I}}}{I}dI'
\eeq
The number of $e^{-}e^{+}$ created by the $\mathcal{E}cos(\omega t)$ field in the $\gamma \ll 1$ and $\gamma \gg 1$ limits, corresponding to the tunneling and multi-photon regimes, respectively, are \cite{ref5,ref6,ref15,ref27,ref74}:
\beq
\begin{split}
& {N(I)} = {2^{ - 3/2}}N_0^4{{\left(\frac{I}{I_c}\right)}^{5/4}}\\
& \times \exp\left[ {-\pi\sqrt{\frac{ I_c}{I}}\left( {1 - {\frac{1}{{2{N_0}^2{{\left( {{\frac{I}{I_c}}} \right)}}}} }} \right)} \right]\left(\frac{\omega \tau }{2\pi}\right)
\quad\mathrm{,}\quad
\gamma \ll 1
\end{split}
\eeq
\beq
{{N(I)} \approx 2\pi N_0^{3/2}{{\left({\frac{8}{{N_0}e}}\right)}^{ - 2{N_0}}}{\left(\frac{I}{I_c}\right)}^{N_0}\left(\frac{\omega \tau }{2\pi}\right)}
\quad\mathrm{,}\quad
\gamma \gg 1
\eeq
where $I$ is the intensity, $e$ is Euler's constant, $I_{c}= 4.65 \times 10^{29}$ W/cm$^{2}$ is the intensity corresponding to $\mathcal{E}_{c}$ and $N_0=\frac{2mc^2}{\hbar \omega}$ is the minimum number photons needed for vacuum pair creation to occur at a given frequency.

Since $\gamma$ is inversely proportional to the square root of the intensity, and the integration in Eqn. (A.2) extends over infinity, we divide the integral into two parts, corresponding to the multi-photon and the tunneling regimes, respectively. The intensity boundary between these two regimes ($I_b$) is set to be the intensity corresponding to $\gamma = 1$. Note that, even if practically such a sharp boundary between the two regimes does not exist, the exact value of this intensity boundary does not need to be well-defined for the purposes of our proof, since when taking the limit of the intensity to zero in Eqn. (A.1), the final result does not depend on its choice.

In view of the above, we write:
\beq
\begin{split}
& \mathop {\lim }\limits_{I \to {0^ + }} \frac{{\tilde N\left( I \right)}}{{N\left( I \right)}}  =  \mathop {\lim }\limits_{I \to {0^ + }} \frac{{2\int_0^\infty  {N\left( {I'} \right)\frac{{{e^{ - \frac{{I'}}{I}}}}}{I}dI'} }}{{N\left( I \right)}}   \\
& = 2\left[ \mathop {\lim }\limits_{I \to {0^ + }} \frac{{\int_0^{{I_b}} {N\left( {I'} \right){e^{ - \frac{{I'}}{I}}}dI'} }}{{N\left( I \right)I}} + \mathop {\lim }\limits_{I \to {0^ + }} \frac{{\int_{{I_b}}^{{\infty}} {N\left( {I'} \right){e^{ - \frac{{I'}}{I}}}dI'} }}{{N\left( I \right)I}} \right]
\end{split}
\eeq
The two terms in the right-hand side of Eqn. (A.5) can now be calculated using Eqns. (A.3) and (A.4) for the number of created pairs in the tunneling and the multi-photon regime, respectively. Note that, in the limit of zero intensities, the number of created papers appearing in the denominator of both terms in Eqn. (A.5) is always given by Eqn. (A.4). By substituting the respective expressions for the number of created pairs of each regime back in Eqn. (A.5), one can show that in the $I \to {0^ + }$ limit, the second term goes to zero. For the first term we have:

\beq
\begin{split}
& \mathop {\lim }\limits_{I \to {0^ + }} \frac{{\int_0^{{I_b}} {N\left( {I'} \right){e^{ - \frac{{I'}}{I}}}dI'} }}{{N\left( I \right)I}} \\
& = \mathop {\lim }\limits_{I \to {0^ + }} \frac{{\int_0^{{I_b}} {{{I'}^{{N_0}}}{e^{ - \frac{{I'}}{I}}}dI'} }}{{{I^{{N_0} + 1}}}} = \mathop {\lim }\limits_{I \to {0^ + }} \left[ { - \Gamma \left( {{N_0} + 1,\frac{{I'}}{I}} \right)} \right]_{{0^ + }}^{{I_b}} \\
& = \mathop {\lim }\limits_{I \to {0^ + }} \left\{ {\mathop {\lim }\limits_{I' \to {0^ + }} \Gamma \left( {{N_0} + 1,\frac{{I'}}{I}} \right) - \Gamma \left( {{N_0} + 1,\frac{{{I_b}}}{I}} \right)} \right\}
\end{split}
\eeq
where $\Gamma(s,x)$ is the upper incomplete gamma function. The first term in the right-hand side of Eqn. (A.6) is
\beq
\begin{split}
& \mathop {\lim }\limits_{I \to {0^ + }} \mathop {\lim }\limits_{I' \to {0^ + }} \Gamma \left( {{N_0} + 1,\frac{{I'}}{I}} \right) = \mathop {\lim }\limits_{I \to {0^ + }} \Gamma \left( {{N_0} + 1,0} \right) \\
& = \mathop {\lim }\limits_{I \to {0^ + }} \Gamma ({{\rm{N}}_0} + 1) = \mathop {\lim }\limits_{I \to {0^ + }} {N_0}! = {N_0}!
\end{split}
\eeq
while for the second one we can show that it approaches zero by using the asymptotic expansion of the upper incomplete gamma function 
\beq
\Gamma (s,z) \sim {z^{s - 1}}{e^{ - z}}\sum\limits_{k = 0} {\frac{{\Gamma (s)}}{{\Gamma (s - k)}}{z^{ - k}}}
\eeq
in the $I \to {0^ + }$ limit. Therefore Eqn. (A.6) becomes
\beq
\mathop {\lim }\limits_{I \to {0^ + }} \frac{{\int_0^{{I_a}} {N\left( {I'} \right){e^{ - \frac{{I'}}{I}}}dI'} }}{{N\left( I \right)I}} = {N_0}!
\eeq
which upon substitution back to Eqn. (A.5) finally proves Eqn. (A.1):
$$ \mathop {\lim }\limits_{I \to {0^+}} \frac{\tilde N(I)}{N(I)} = 2N_0! $$

\end{document}